# International R&D Spillovers and other Unobserved Common Spillovers and Shocks[*]

Diego-Ivan Ruge-Leiva[**]

January 2015

.

**Abstract**

Studies which are based on Coe and Helpman (1995) and use weighted foreign R&D variables to estimate channel-specific R&D spillovers disregard the interaction between international R&D spillovers and other unobserved common spillovers and shocks. Using a panel of 50 economies from 1970-2011, we find that disregarding this interaction leads to inconsistent estimates whenever knowledge spillovers and other unobserved effects are correlated with foreign and domestic R&D. When this interaction is modeled, estimates are consistent; however, they confound foreign and domestic R&D effects with unobserved effects. Thus, the coefficient of a weighted foreign R&D variable cannot capture genuine channel-specific R&D spillovers.

***A supplement to this article is available online at:

http://mpra.ub.uni-muenchen.de/62205/1/MPRA_paper_62205.pdf

[*] I would like to thank Ron Smith, Bang Jeon, Tomás Mancha Navarro, David Arturo Rodriguez, Camilo Andres Mesa and Juan Manuel Santiago for helpful comments and suggestions. I also thank Robert Inklaar for answering my questions about the Penn World Tables 8.0, and Alvaro Garcia Marin for sharing the data set of Lederman and Saenz (2005). I thank Markus Eberhardt for sharing his econometric routines with me and for his valuable comments and support during my studies at the University of Nottingham, UK, which has been fundamental for the empirical analysis of this study. I am especially grateful to German Umana and Xueheng Li for their unconditional guidance and encouragement. The author also wishes to acknowledge the financial support of the Economics Department of the Universidad Central, Colombia.
[**] Universidad Central, Colombia. Economics Department. Email: drugel@ucentral.edu.co

# 1. Introduction

In the past three decades there has been a great deal of research into the estimation of the empirical significance of international R&D spillovers at the country level. A large number of these studies are mainly based on the endogenous economic growth theory, which states that technological development and productivity growth can be achieved by the spread of technology through international trade driven by profit-seeking firms, that is, a situation where the recipient countries employ technology as an intermediate input in order to develop a larger range of inputs or inputs of a higher quality (Romer, 1990; Grossman and Helpman, 1991; Aghion and Howitt, 1992). International R&D spillovers therefore occur when investment in the development of new inputs increases the levels of R&D investment and reduces future R&D costs across nations, and today's improvement of the available domestic and foreign products allows future innovators to improve the quality of these products, insofar as they can do that at a faster rate when the initial quality of such products is higher, which, in turn, increases the productivity of intermediate inputs, such as R&D (Coe et al. 2009).

The first empirical study which applied these theoretical concepts was done by Coe and Helpman (1995) (hereafter CH). In it, they investigate how countries may benefit from imports, in accordance with the technological knowledge of their trade partners and their own degree of openness. Towards that end, CH introduce a domestic and a weighted foreign R&D capital stock variables in a Total Factor Productivity (TFP) function,[1] in a way that the country-specific foreign R&D capital stock measure takes into account trade-based technology transfers from all the countries in the sample. This measurement is therefore based on the weighted average of the domestic R&D from country partners where bilateral imports are used as weights. CH find, first, that knowledge spillovers and returns to domestic R&D, which are estimated through the coefficient of the foreign and domestic R&D variables respectively, are statistically significant in determining cross-country productivity; second, that the more open the economy, the larger the effect of knowledge spillovers; and third, that the returns to domestic R&D are larger for the G7 countries, whereas knowledge spillovers are larger for the smaller advanced economies.

Other empirical studies, which follow the CH framework but employ channels of knowledge diffusion different from trade and/or use different weighting schemes for the foreign R&D variable, likewise claim that returns to domestic R&D and international R&D spillovers explain productivity and can be accurately estimated through the coefficients of variables for domestic and weighted foreign R&D, respectively.[2]

Two assumptions at the core of these empirical studies support their conclusions: first, that the CH framework assumes error cross-section independence, which implies that the interplay between international R&D spillovers and other unobserved common spillovers and shocks does not cause contemporaneous correlation across countries;[3] and second, a weighted foreign R&D variable is imposed in order to only detect channel-specific R&D spillovers. This effect is assumed not to arise from the interaction between unobserved spillovers and shocks, whose impact is uncorrelated with R&D and productivity, but merely from this weighted variable.

We, on the other hand, would argue that in any economic environment, the R&D spillovers which spread through a specific channel and which are unobserved may be mixed with knowledge spillovers transferred through other channels, along with other unobserved micro and macroeconomic spillovers and shocks which are associated with productivity and R&D. We therefore assert that the abovementioned spillover variable does not sufficiently address this interaction between R&D spillovers and other unobservable effects. That is because it is assumed that its coefficient successfully captures genuine channel-specific R&D spillovers, without clarifying how this sort of variable could separate this effect from other unobserved common effects. In such a situation, the estimate of a foreign R&D variable might represent other aspects.

---

[1] In the present paper, the term "total factor productivity" is equivalent to "productivity."
[2] A brief review of this literature is discussed in section 1 of the online supplement.
[3] Hereafter I will use the terms "unobserved common spillovers and shocks," "unobserved common factors," "unobserved common effects," "unobservables" and similar words interchangeably.

Furthermore, if this variable is employed without regarding the interaction between R&D spillovers and other unobserved effects, and if all these unobservables are correlated with the variables of the model as sources of cross-section dependence, then the consistency of the foreign and domestic R&D estimates could be affected. In fact, even if the interplay between unobserved effects is taken into account, the spillover variable will not necessarily serve to capture genuine R&D spillovers.

In order to study these concerns, the present article contributes to the existing literature on international R&D spillovers according to the following features: First, we study the empirical results of introducing a weighted foreign R&D variable in the CH framework without accounting for the interaction between international R&D spillovers diffused by any channel and other unobserved heterogeneous spillovers and shocks, which are common across countries, may jointly occur as sources of cross-section dependence and might be correlated with the variables of the model. Second, we examine the estimates of the domestic and weighted foreign R&D variables in a multifactor error structure where we regard the interaction between international knowledge spillovers and other weak and/or strong unobservables detected in the error term, and compare its results with those of the CH approach.

Third, we employ several estimators in static and dynamic models to study the long-term effects of the R&D variables on productivity according to the CH approach and the multifactor framework, although we mainly rely on the results of the set of dynamic models that account for unobservables, because they can be regarded as complementary when dealing with several econometric issues, which we document in the paper. We use a weighted foreign R&D variable in line with Lichtenberg and van Pottelsberghe de la Potterie (1998, hereafter LP), which will account for knowledge transmission through trade from all countries of the sample.[4] Fourth, for the purpose of gauging the reliability of the estimates at the aggregate level, this study allows technology parameters to differ across countries. It employs a sample of 50 emerging and advanced economies from 1970-2011 which explains several contemporary, heterogeneous cross-country interdependencies.

Our results suggest that first, trade-related R&D spillovers cannot be estimated through the coefficient of an imposed spillover variable in the CH approach. This is because introducing this variable while ignoring the interaction between unobservables, which may be correlated with the covariates, leads to seriously biased and inconsistent estimates. Second, when the interplay between international R&D spillovers spread by any channel and other unobserved effects is regarded in a multifactor error structure, significant foreign and domestic R&D estimates become consistent and not seriously biased in most cases. However, most of them are larger than those from a CH specification, since they are subject to weak residual cross-section dependence, which indicates that the estimates are capturing the effect of unobservables in addition to the direct effect of the R&D variables.

---

[4] We present additional results in an online supplement using both a LP and a CH weighted foreign R&D variable with several weighting configurations.

In this case trade-related knowledge spillovers cannot be identified. Therefore, nothing ensures that the coefficient of a spillover variable captures genuine R&D spillovers. Instead, it might be capturing other data cross-section dependencies. Moreover, contrary to the CH approach, returns to R&D are not independent of their associated spillovers and shocks. In fact, domestic R&D estimates can also be affected by the unobserved effects associated with the weighted foreign R&D variable.

These findings are of a crucial relevance for developing countries because they indicate that the identification and measurement of the international R&D spillovers spread by trade or any other specific channel must be done in a more suitable empirical framework where we can account for the interplay between this effect, R&D spillovers transferred by other channels and other unobserved common spillovers and shocks that may be sources of error cross-section dependence. Otherwise, empirical studies may yield inaccurate information for economic analysis and R&D policies of developing countries.

Studies by Belitz and Molders (2013) and Ertur and Musolesi (2013) have analyzed the effect of the domestic and foreign R&D (weighted by different schemes) on productivity in order to account for unobservables in a multifactor error structure. However, these studies neither address the abovementioned issues nor discuss the importance of identifying channel-specific R&D spillovers when regarding other unobserved effects.

In its empirical spirit the present study is closest to that of Eberhardt et al. (2013), which deals with some of the above issues. They analyze the effect of R&D on value added for 12 industries across 10 advanced countries, accounting for unobservables, and find that the approach of Griliches (1979), which ignores unobserved effects, yields sizable and significant R&D estimates, although it is misspecified due to residual cross-section dependence. When unobservables are regarded in a multifactor error structure, the R&D estimates are consistent, but fall in magnitude and significance. This evidence shows that R&D and spillovers are indivisible when estimating private returns to R&D.

Further, Eberhardt et al. (2013) claim to find that weighted R&D variables capture broader cross-sectional dependence than solely genuine R&D spillovers. However, they do not provide any empirical evidence to show the circumstances under which this occurs, and do not include this sort of variable in any of their empirical specifications to analyze its coefficient. They only estimate the effect of domestic R&D stock, labor and capital stock on value added by regarding the presence of unobservables in a multifactor framework, and claim this approach is more appropriate than using a spillover variable to model unobserved effects. By contrast, our estimates at the country level of a weighted foreign R&D variable when unobservables have been accounted for clearly demonstrate that they capture the effect of several sources of cross-section dependence than only pure R&D spillovers.

The rest of this paper is organized as follows: Section 2 reviews the Coe and Helpman model. Section 3 introduces a multifactor error structure for international R&D spillovers and other unobserved common effects. Section 4 presents the estimation methodology. Section 5 describes the data and introduces a cross-section dependence and unit root tests. Section 6 shows the results, which we discuss in Section 7. Section 8 presents our conclusions.

**2. The Coe and Helpman Model**

The simplest empirical model proposed by CH, which is based on the theories of innovation-driven endogenous technological change, can be written as follows:

$$tfp_{it} = \alpha_i + \beta_{i1} R_{it}^d + \beta_{i2} R_{it}^f + u_{it} \qquad (1)$$

where $tfp_{it}$, $R_{it}^d$ and $R_{it}^f$ are the logarithmic variables of total factor productivity,[5] domestic R&D capital stock and foreign R&D capital stock respectively. These regressors are specific to country $i$ at time $t$ for $i = 1, \ldots, N$, and $t = 1, \ldots, T$. $\alpha_i$ is a constant term which accounts for country-specific effects. We include the error term $u_{it}$ in (1) according to a panel data setting.

According to (1), domestic R&D contributes to the availability and/or quality of inputs of a country, while foreign R&D represents the R&D capital stock from the rest of the world which is available for a specific country through international trade. More importantly, international R&D spillovers, which are spread through trade, are assumed to be captured by the coefficient of the foreign R&D $\beta_{i2}$, because they arise from the transmission of the R&D capital stock by bilateral trade from foreign countries to the home country $i$. The coefficient of the domestic R&D $\beta_{i1}$ is assumed to show the contribution of domestic R&D, separately from knowledge spillovers.

As stated by CH, a convenient way to represent the foreign R&D capital stock is to aggregate the R&D capital stocks of foreign countries in $R_{it}^f$ as follows:

$$R_{it}^f = \sum_{i \neq c} w_{ic,t} R_{ct}^d \qquad (2)$$

where $w_{ic,t}$ are weights of the cumulative R&D expenditures of the country $i$'s foreign trading partners $c$, which are defined by bilateral imports and are allowed to vary over time. This specification points to the fact that the domestic economy will benefit more from the international knowledge spillovers which arise from bilateral trade when the domestic R&D of their partners is large.

The weighting scheme suggested by CH is the import-share-weighted average of the domestic R&D capital stock of trade partners.[6] Some studies on international knowledge spillovers at the country level have suggested alternative weighting schemes, such as bilateral imports multiplied by the R&D/GDP ratio of foreign countries (LP), technological proximity (Guellec and van Pottelsberghe de la Potterie, 2004), technological proximity and shares of patent citations (Lee, 2006), and equal weights (Keller, 1998). They have likewise proposed other channels of transmission of R&D, such as inward and outward FDI (van Pottelsberghe de la Potterie and Lichtenberg, 2001), exports (Funk, 2001), migration of students (Park, 2004), and the transfer of information technology (Zhu and Jeon, 2007).

---

[5] CH define TFP as $\log Y - \xi \log K - (1 - \xi) \log L$, where $Y$ is the GDP, $K$ the capital stock, $L$ the available labor force, and $\xi$ is the share of capital in GDP. In the context of the present article, however, TFP is defined differently (see the appendix for more details).

[6] CH argue that equation (1) may not capture the role of international trade because the weights are fractions that add up to one so that they do not properly show the level of imports. Therefore, they propose another model where they multiply the foreign R&D variable and the level of imports as a measure of openness. However, in the present paper an openness variable does not interact with the foreign R&D variable. Instead, I follow the basic framework found in the literature because this will be sufficient to show the implications of disregarding the interaction between R&D spillovers and other unobserved spillovers and shocks in the CH specification.

It is worth noting that two additional assumptions characterize the CH approach: first, the model (1) for the analysis of international R&D diffusion is subject to error cross-section independence, that is, there are no contemporary interdependencies across countries caused by the interaction between international R&D spillovers spread by any channel and other unobserved spillovers and shocks which are detected in the error term; and second, a spillover variable, such as that defined in (2), is imposed in equation (1) in order to capture R&D spillovers which spread across borders through only one channel, such as trade. It is assumed that these spillovers do not arise from unobservables, which remain neutral to TFP and R&D, but solely arise from the spillover variable.

From our standpoint, these assumptions might be restrictive when dealing with unobserved effects. This is because a spillover variable may not take account of the fact that international R&D spillovers which spread through a specific channel (trade in the case of CH) are likely to arise together with a variety of other unobserved common effects, such as R&D spillovers which are transmitted through other channels, pecuniary R&D spillovers,[7] linkage and measurement spillovers and, in general, micro and macroeconomic spillovers and shocks which may be correlated with TFP and R&D. Moreover, it is not clear how the coefficient of this variable can distinguish between channel-specific R&D spillovers and other unobservables. To first assume cross-section independence and then impose a spillover variable might not be appropriate for estimating channel-specific knowledge spillovers. Under these circumstances, the coefficient of a spillover variable might be capturing other effects than R&D spillovers.

We further believe that even if a spillover variable is incorporated alongside a domestic R&D variable in the CH framework, the consistency of the foreign and domestic R&D estimates might be seriously affected if the interplay between unobserved effects, which may be correlated with the variables of the model, is not properly taken into account as a source of error cross-section dependence. In fact, if we account for international R&D spillovers spread by any channel and other weak and/or strong unobservables, which are sources of cross-section dependence that might arise together, nothing would ensure that estimates of a spillover variable can capture genuine R&D spillovers even if the estimates of the model are consistent.

To explain the above, we will now show how the interaction between unobserved effects may bring about error cross-section dependence.

## 3. The Multifactor Error Structure for International R&D Spillovers and Other Unobserved Common Spillovers and Shocks

According to Pesaran (2006), Chudik et al. (2011), Chudik and Pesaran (2013a) and other recent investigations on macroeconometric panel time series models, one of the ways to deal with the error cross-section dependence caused by unobservables is to use a multifactor error structure in which sources of cross-section dependence are assumed to be represented by a few unobserved common factors that affect all the observations and can be found in the error term. Applying this approach, we can therefore write an extension of (1) which accounts for international R&D spillovers and other unobserved spillovers and shocks which might arise together:

$$tfp_{it} = \beta_1 R_{it}^d + \beta_2 R_{it}^f + u_{it}, \quad u_{it} = \gamma_{i1} f_{1t} + \cdots + \gamma_{im} f_{mt} + \varepsilon_{it} = \boldsymbol{\gamma}_i' \boldsymbol{f}_t + \varepsilon_{it} \qquad (3)$$

---

[7] According to Hall et al. (2009), R&D pecuniary spillovers arise through transactions between firms which produce new or improved intermediate goods at prices which reflect less than the total value of the progress incorporated.

where each $f_{jt}$, for $j = 1, ..., m$, is a single unobserved common factor that affects all cross-sectional units, although in different degrees, depending on the magnitude of its $j^{th}$ heterogeneous factor loading, $\gamma_{ij}$. $\boldsymbol{\gamma}_i$ is a $m \times 1$ vector of factor loadings, and $\boldsymbol{f}_t$ a $m \times 1$ vector of unobserved common factors. $\varepsilon_{it}$ are the idiosyncratic errors.

Factors $f_{jt}$ represent two categories of shocks and spillovers: (i) at the macroeconomic level, such as aggregate financial shocks, real shocks, global R&D and technology spillovers, or structural changes; and (ii),at the microeconomic level, such as local spillovers which arise from industrial activity and domestic technology development, local consumption and income effects, socioeconomic networks, and geographic proximity. Among the examples of positive and negative unobserved common shocks and spillovers in the time frame we analyze there are international R&D spillovers which spread through any bilateral or multilateral channel (such as trade, FDI, or migration), the oil crisis of the 1970s, the financial crisis in Latin America during the 1980s, the standardization of the Internet Protocol Suite (TCP/IP), the downfall of communism, the global financial crisis of 2008, and the emergence of China and India as major global economies during the 21th century

Such spillovers and shocks are common because they affect all countries, although their impact is heterogeneous. In extreme cases, they may either affect all countries with a strong heterogeneous impact, or have a weak effect (or no effect at all) on a subset of countries. Observed common factors (such as the prices of commodities) or deterministics (intercepts or seasonal dummies) are omitted in (3) for the purpose of brevity (i.e. $\alpha_i = 0$), even though they may be easily included. Now, when we place the factors in $u_{it}$ into the $tfp_{it}$ function in (3), it yields the extended model:

$$tfp_{it} = \beta_{i1} R_{it}^d + \beta_{i2} R_{it}^f + \gamma_{i1} f_{1t} + \cdots + \gamma_{im} f_{mt} + \varepsilon_{it} \qquad (4)$$

or more compactly:

$$tfp_{it} = \beta_{i1} R_{it}^d + \beta_{i2} R_{it}^f + \boldsymbol{\gamma}_i' \boldsymbol{f}_t + \varepsilon_{it} \qquad (5)$$

It can easily be seen from (4) and (5) that shocks and spillovers are now present as unobserved factors that determine TFP. Now let us show the possible consequences of introducing a spillover variable $R_{it}^f$ in this framework by allowing the correlation between the individual specific regressors, $R_{it}^d$ and $R_{it}^f$, and the error term $u_{it}$, on the assumption that the first two can be determined by the impact of their associated unobserved factors:

$$R_{it}^d = \Gamma^d{}_{i1} f_{1t} + \cdots + \Gamma^d{}_{im} f_{mt} + v_{it} = \boldsymbol{\Gamma}_i^{d'} \boldsymbol{f}_t + v_{it} \qquad (6)$$

$$R_{it}^f = \Gamma^f{}_{i1} f_{1t} + \cdots + \Gamma^f{}_{im} f_{mt} + s_{it} = \boldsymbol{\Gamma}_i^{f'} \boldsymbol{f}_t + s_{it} \qquad (7)$$

where factor loadings $\Gamma^d$ and $\Gamma^f$ represent the magnitude at which factors are correlated with $R_{it}^d$ and $R_{it}^f$, respectively. $\boldsymbol{\Gamma}_i^d$ and $\boldsymbol{\Gamma}_i^f$ are $m \times 1$ invertible matrices of factor loadings, and $v_{it}$ and $s_{it}$ are idiosyncratic components of $R_{it}^d$ and $R_{it}^f$, respectively, which are assumed to be distributed independently of the innovations $\varepsilon_{it}$.

Now, if we add (6) and (7), define the result in terms of the shocks $\boldsymbol{f}_t$, and introduce this into (5) where we can factorize $R_{it}^d$ and $R_{it}^f$, we obtain:

$$tfp_{it} = \left[\beta_{i1} + \boldsymbol{\gamma}_i'\left(\boldsymbol{\Gamma}_i^{d'} + \boldsymbol{\Gamma}_i^{f'}\right)^{-1}\right]R_{it}^d + \left[\beta_{i2} + \boldsymbol{\gamma}_i'\left(\boldsymbol{\Gamma}_i^{d'} + \boldsymbol{\Gamma}_i^{f'}\right)^{-1}\right]R_{it}^f + \phi_{it}, \qquad (8)$$

where the coefficients of $R_{it}^d$ and $R_{it}^f$ are subject to the magnitude of the impact of the unobserved common effects, and where $\phi_{it} = -\boldsymbol{\gamma}_i'\left(\boldsymbol{\Gamma}_i^{d'} + \boldsymbol{\Gamma}_i^{f'}\right)^{-1}(v_{it} + s_{it}) + \varepsilon_{it}$. From (8) we can see, first, that the coefficient of the foreign R&D variable $R_{it}^f$ confounds the effect of this variable $\beta_{i2}$, and that of international R&D spillovers and other weak and/or strong unobserved common spillovers and shocks, represented by $\boldsymbol{\gamma}_i'\left(\boldsymbol{\Gamma}_i^{d'} + \boldsymbol{\Gamma}_i^{f'}\right)^{-1}$. This shows that when the effect of a mixture of unobservables is accounted for, channel-specific knowledge spillovers cannot be identified through the coefficient of a spillover variable. Second, the coefficient of the domestic R&D variable represents a combination of returns to domestic R&D $\beta_{i1}$ and the effect of shocks and spillovers associated with the domestic R&D regressor through $\boldsymbol{\Gamma}_i^{d'}$, which shows that the two might not be separate. However, the introduction of the weighted foreign R&D variable and the effect of its associated shocks $\boldsymbol{\Gamma}_i^{f'}$ could affect the coefficient of the domestic R&D regressor and therefore the results of the model.

Based on Chudik et al. (2011), we represent the magnitude of the impact of shocks through the factor loadings as follows:

$$\lim_{N\to\infty} N^{-\alpha} \sum_{i=1}^{N} |\gamma_{ij}| = K < \infty \qquad (9)$$

where $K$ is a fixed positive constant that does not depend on the number of countries, $N$. Given (9), factors are said to be weak if $\alpha = 0$, semi-weak if $0 < \alpha < 1/2$, and semi-strong if $1/2 < \alpha < 1$. For these sorts of factors we can say that the multifactor error structure is cross-sectionally weakly dependent at a given point in time $t \in T$, where $T$ is an ordered time set, if $\alpha < 1$. In this case, weak, semi-weak and semi-strong factors may produce estimates of the domestic and foreign R&D which are not seriously biased and whose consistency and asymptotic normality are not affected. These factors may only affect a subset of countries of the whole sample and the number of affected economies rises less than the total countries of the sample.

On the other hand, factors are strong if $\alpha = 1$ in (9), so that the multifactor error structure is cross-sectionally strongly dependent at a given point in time $t \in T$ if and only if there exists at least one strong factor.[8] In that case, it is possible that the factors may be correlated with the domestic and foreign R&D, in such a way that the models yield seriously biased and inconsistent estimates. Chudik and Pesaran (2013b) characterize the strong factors as those which reflect the pervasive effect of error cross-section dependence in the sense that they affect all countries in the sample and their effect is persistent even if $N$ tends to infinite. Furthermore, if unobserved weak and strong common factors are disregarded, and if these factors are correlated with the variables of the model, then the consistency of the estimates may also be severely affected.

---

[8] According to Chudik and Pesaran (2013b) the overall exponent $\alpha$ can be defined as $\alpha = max(\alpha_1, \dots, \alpha_m)$.

## 4. Estimation Methodology

In order to address the above concerns, we employ a variety of estimators for the CH model defined in (1), which ignores unobservables, and the multifactor error structure in (4), which accounts for unobserved common effects (including an intercept, i.e. $\alpha_i \neq 0$). This estimation strategy helps us to analyze the coefficients of the domestic R&D and weighted foreign R&D variables under different empirical assumptions, and provides useful information for a comparison of the results of different empirical approaches.

The first set of estimators is used in static models on the following assumptions: first, the estimators restrict homogeneity in the technology parameters and (i) assume error cross-section independence (in line with CH), such as pooled OLS (POLS), first difference (FD), and two-way fixed effects (2FE); or (ii) allow for error cross-section dependence (i.e. account for unobservables), such as the Pesaran (2006) Common Correlated Effects (CCE) pooled estimator (CCEP) with strictly exogenous regressors.[9] Second, estimators which allow for the technological heterogeneity of slopes and (i) assume error cross-section independence such as the mean group (MG) estimator and the cross-sectionally demeaned MG (CDMG) estimator; or (ii), allow for error cross-section dependence such as the heterogeneous CCE (CCEMG) with strictly exogenous regressors.[10] In contrast to the estimators which disregard unobserved spillovers and shocks, the CCE approach includes cross-section averages of variables in a common factor framework as proxies for unobserved common factors,[11] so long as the weights of these averages satisfy certain granularity and normalization conditions.[12]

---

[9] In accordance with Engle et al. (1983), a process that is weakly exogenous is characterized by (i) a reparametrization of the parameters of interest and (ii) a (classical) sequential cut condition. This validates the idea of making inference conditional on the regressors; however, it is worth noting that Granger causal feedback effects may implicitly arise at some point. A process that is strictly exogenous, on the other hand, is characterized by weak exogeneity plus Granger noncausality from a dependent variable onto the regressors (the latter is essential to validate forecasting the independent variables and then forecast the dependent variable conditional on leads of regressors), i.e. there are no Granger causal feedbacks.

[10] Even though we account for the impact of the interplay between unobserved common effects using the CCE estimator, this approach does not allow us to study the specific nature of each of those unobserved effects. For an accurate estimate of channel-specific R&D spillovers, more research on this aspect needs to be done.

[11] This is because cross-section averages pool information on markets, i.e. they pool the past and current views of economic agents on the constitution of covariates. Further, Pesaran and Tosetti (2011) state that the effects of temporal and spatial correlations due to spatial and/or unobserved common factors are eliminated by the addition of cross-section averages.

[12] The CCE approach to static models has several econometric advantages. First, it does not require prior knowledge of the number of unobserved common factors (Pesaran 2006); second, CCE estimates are consistent even when there is serial correlation in errors (Coakley et al. 2006); third, it is consistent and asymptotically normal when the idiosyncratic errors are characterized by a spatial process (Pesaran and Tosetti 2011) and when errors are subject to a finite number of unobserved strong effects and an infinite number of weak and/or semi-strong unobserved common effects so long as that certain conditions on the factor loadings are satisfied

In this case, (4) becomes:

$$tfp_{it} = \alpha_i + \beta_{i1}R_{it}^d + \beta_{i2}R_{it}^f + \boldsymbol{\psi}_i'\bar{\mathbf{z}}_t + \varepsilon_{it} \qquad (10)$$

where $\bar{\mathbf{z}}_t = (\overline{tfp}_t, \bar{\mathbf{x}}_t')'$ are the cross-section averages of the TFP and the domestic and foreign R&D variables, which are represented by $\mathbf{x}_{it} = (R_{it}^d, R_{it}^f)'$.

We also apply our empirical analysis to dynamic models by using a second set of estimators. Three models are employed in this case (the first two in an ECM representation) where we estimate the long-run effects of the domestic and foreign R&D variables on TFP:[13] (i) the traditional autoregressive distributed lag (ARDL), (ii) the cross-sectional ARDL (CS-ARDL) with heterogeneous technology parameters and the weakly exogenous regressors (aka dynamic CCEMG) found in Chudik and Pesaran (2013a); and (iii), the heterogeneous cross-sectional distributed lag (CS-DLMG) approach of Chudik et al. (2013), which does not include lags of the dependent variable. The first model is the traditional ARDL approach, which is used to obtain the long-run estimates of the domestic and foreign R&D variables in a dynamic setup of the CH framework. The model is defined as follows:

$$tfp_{it} = \alpha_i + \sum_{l=1}^{p} \varphi_{il} tpf_{i,t-l} + \sum_{l=0}^{p} \boldsymbol{\beta}_{il}' \mathbf{x}_{i,t-l} + u_{it} \qquad (11)$$

where $\boldsymbol{\beta}_{io}' = (\beta_{i1,0}, \beta_{i2,0})$ for $l = 0$ in (11), in accordance with the coefficients of the domestic and foreign R&D in (10). $p = 1$ to 3 lags are considered for the ARDL model in order to include sufficiently long lags, given the time period of the sample, and to fully account for the short-run dynamics and thus derive the long-term coefficients, assuming that there is a single long-run relation between the dependent variable and the independent variables.[14] The ARDL model in (11) can also be written in an ECM representation, as follows:

$$\Delta tfp_{it} = \alpha_i - \lambda_i (tfp_{i,t-1} - \boldsymbol{\theta}_i \mathbf{x}_{i,t-1}) + \sum_{l=1}^{p-1} \phi_{il} \Delta tfp_{i,t-l} + \sum_{l=0}^{p-1} \boldsymbol{\pi}_{il}' \Delta \mathbf{x}_{i,t-l} + u_{it} \qquad (12)$$

where $\Delta tfp_{it} = tfp_{it} - tfp_{it-1}$, $\Delta \mathbf{x}_{it} = \mathbf{x}_{it} - \mathbf{x}_{it-1}$, $\lambda_i = 1 - \sum_{l=1}^{p} \varphi_{il}$, $\phi_{il} = -\sum_{k=l+1}^{p} \varphi_{ik}$ for $1 \leq l \leq p-1$, $\boldsymbol{\theta}_i = (\sum_{l=1}^{p} \boldsymbol{\beta}_{il})/\lambda_i$, $\boldsymbol{\pi}_{i0} = \boldsymbol{\beta}_{i0}$, and $\boldsymbol{\pi}_{il} = -\sum_{k=l+1}^{p} \boldsymbol{\beta}_{ik}$ for $1 \leq l \leq p-1$. Estimations are carried out according to (12) by employing the POLS, 2FE and MG estimators (all estimators assume error cross-section independence in line with CH).

---

(Chudik et al. 2011); fourth, the CCE estimator with either stationary or nonstationary factors has a similar asymptotic distribution when they are cointegrated, and even the latter could be noncointegrated (Kapetanios et al. 2011); and fifth, it can be extended to unbalanced panels (Chudik and Pesaran 2013b).

[13] Short-run estimates will be available upon request.

[14] In this model we assume that lags are the same across variables and countries because, as stated in Chudik et al. (2013), this helps to reduce the adverse effects of the selection of data which may be subject to the use of lag order selection procedures, such as the Akaike or Schwarz criteria.

As reported by Chudik et al. (2013), the ARDL structure is valid regardless of whether the independent variables are exogenous or endogenous, or characterized as order one, I(1), or order zero, I(0), processes. In fact, long-term estimates of $\boldsymbol{\theta}_i$ (which can be arrived at through the estimates of the short-term coefficients $\boldsymbol{\beta}_{il}$ and $\varphi_{il}$) may be consistent when common factors are serially uncorrelated and when they are uncorrelated with the regressors. This favors consistent estimation, especially when there is reverse causality, i.e. when past values for productivity may determine current domestic and foreign R&D capital stocks. We can also state that the ARDL models in an ECM representation are convenient because we can estimate the mean of the coefficients of the error correction term, denoted by $\lambda_{il}$, in order to study the long-run cointegration of covariates as well as analyze the speed of convergence towards the long-term equilibrium of steady state.

It is worth noting that this approach has some drawbacks. First, there could be a large sampling uncertainty due to the restricted time dimension of the panel and the slow speed of convergence towards the long-term. Second, as Pesaran and Smith (1995) have shown, under a random coefficient model which characterizes heterogeneous dynamic panel data models, pooled OLS estimators are no longer consistent. Third, the ARDL model requires an appropriate choice of lag orders to obtain proper long-run estimates.

The second model is the CS-ARDL approach which allows for heterogeneous technology coefficients and, in contrast to the traditional ARDL model, includes cross-section averages in a dynamic multifactor framework as proxies for R&D spillovers spread by any channel and other unobservables. The CS-ARDL models can be expressed as follows:

$$tfp_{it} = \alpha_i + \sum_{l=1}^{p} \varphi_{il} tpf_{i,t-l} + \sum_{l=0}^{p} \boldsymbol{\beta}'_{il} \boldsymbol{x}_{i,t-l} + \sum_{l=0}^{3} \boldsymbol{\psi}'_{il} \bar{\boldsymbol{z}}_{t-l} + e_{it} \quad (13)$$

where $\bar{\boldsymbol{z}}_{t-l} = (\overline{tfp}_{t-l}, \bar{\boldsymbol{x}}'_{t-l})'$ are the contemporaneous and lagged cross-section averages of the dependent and independent variables, which are chosen on the basis of the rule of thumb $T^{1/3}$. The present study allows for up to $T^{1/3} = 41^{1/3} \approx 3$ lagged cross-section averages of each variable, independently of the number of the lags of the variables of (13), for which $p = 1, 2$ and 3 lags are included where a maximum number of unobserved factors (which might be small) is assumed. $e_{it}$ is determined by Chudik and Pesaran (2013b) on the basis of three aspects: (i) an idiosyncratic term $\varepsilon_{it}$, (ii) an error component due to the approximation of unobserved common factors based on large $N$ relationships, and (iii) an error component that is explained by the truncation of a possibly infinite polynomial distributed lag function. As can be seen, lagged cross-section averages allow for the possibility that unobserved common spillovers and shocks may react to lags. An ECM representation of this model can be easily written as follows:

$$\Delta tfp_{it} = \alpha_i - \lambda_i \big(tfp_{i,t-1} - \boldsymbol{\theta}_i \boldsymbol{x}_{i,t-1}\big) + \sum_{l=1}^{p-1} \phi_{il} \Delta tfp_{i,t-l} + \sum_{l=0}^{p-1} \boldsymbol{\pi}'_{il} \Delta \boldsymbol{x}_{i,t-l} + \sum_{l=0}^{3} \boldsymbol{\psi}'_{il} \bar{\boldsymbol{z}}_{t-l} + e_{it} \quad (14)$$

However, this approach has been formulated only for stationary panels and is subject to sampling uncertainty when the time period is not large enough.

The third dynamic panel data model is the CS-DLMG approach proposed by Chudik et al. (2013), which allows for the heterogeneity of technology coefficients and unobserved effects. This approach can be obtained by subtracting $\sum_{l=1}^{p} \varphi_{il} tpf_{i,t-l}$ from both sides of (13), factorizing $A_i(L) = (1 - \sum_{l=1}^{p} \varphi_{il} L^l)$ and then dividing the whole equation by this expression, in order to arrive at the following equation:

$$tfp_{it} = \varrho_i + \dot{\boldsymbol{\theta}}_i' \boldsymbol{x}_{it} + \sum_{l=0}^{p-1} \boldsymbol{\delta}_{il}' \Delta \boldsymbol{x}_{i,t-l} + \omega_{i,tfp} \overline{tfp}_t + \sum_{l=0}^{3} \boldsymbol{\omega}_{i,xl}' \overline{\boldsymbol{x}}_{t-l} + \tilde{e}_{it} \qquad (15)$$

where $\tilde{e}_{it} = A_i(L)^{-1} e_{it}$, $\varrho_i = A_i(L)^{-1} \alpha_i$, $\dot{\boldsymbol{\theta}}_i = A_i(L)^{-1} \sum_{l=0}^{p} \boldsymbol{\beta}_{il}$, $\boldsymbol{\delta}_{il} = A_i(L)^{-1} \boldsymbol{\varsigma}_{il}$, $\boldsymbol{\varsigma}_{il} = -\sum_{k=l+1}^{p} \boldsymbol{\beta}_{ik}$ for $0 \leq l \leq p-1$, and $\overline{tfp}_t$ and $\overline{\boldsymbol{x}}_t$ are cross-section averages of TFP and the R&D variables respectively, where we allow for lagged cross-section averages of $\overline{\boldsymbol{x}}_t$ only. The loadings $\omega_{i,tfp}$ and $\boldsymbol{\omega}_{i,xl}$ are different from those of (13) (such as $\boldsymbol{\psi}_{il}$) because they contain $A_i(L)^{-1}$. Here, the CS-DLMG models are estimated by adding three lagged cross-section averages. The present study takes advantage of the fact that $\dot{\boldsymbol{\theta}}_i$ can be consistently estimated directly by the CCE approach, and only requires a selection of a truncation lag, in contrast to the ARDL approach, which depends on a correct specification of the lags order. In addition, the $p = 1, 2$ and $3$ lags of the regressors are included.

Once cross-section averages are included in the model, it is possible to obtain robust estimates even when the time period is short. They are also robust to the presence of nonstationary variables and factors (regardless of the number of unobserved factors), weak cross-section dependence, serial correlation or breaks in the idiosyncratic errors and serial correlation in unobserved factors. However, the CS-DLMG does not properly tackle the problem of the feedback effects from lagged values of the TFP on the domestic and foreign R&D, so long-term estimates are consistent only in the absence of this problem. Furthermore, estimates for small samples are only consistent so long as the roots of $A_i(L)$ fall strictly outside the unit circle.[15]

The present study has followed Chudik et al. (2013) in the sense that we use the CS-DLMG and the CS-ARDL estimators as complementary when dealing with several econometric questions and to obtain robust results. However, we mainly rely on the results of the CS-ARDL model in an ECM specification, because the cointegration of variables in the long-term can be easily observed and this model deals with a variety of problems which characterize R&D capital stock and unobserved common effects: the lagged effects of R&D and unobserved spillovers and shocks associated with the TFP and R&D variables, and the feedback effects of past productivity values on the R&D regressors.

## 5. Data, and Cross-Section Dependence and Unit Root Tests

The data set contains aggregate data from 1970 to 2011 for 50 advanced and emerging countries for an unbalanced panel with $N_{min} = 20$ and $T_{min} = 20$. The data sources and the methodologies employed to construct the variables are included in the Appendix. Information on the data set is reported in Table 1. There are 2042 observations for the TFP, 1873 for the domestic R&D capital stock and 2056 for the LP weighted foreign R&D capital stock, whose weights allow for knowledge transmission through trade from all the countries of the sample.

---
[15] In this case, the coefficients $\boldsymbol{\delta}_{il}$ are exponentially decaying due to $A_i(L)$.

<<INSERT TABLE 1 HERE>>

An online supplement to the present paper includes: first, plots of all the series; second, Stata routines; third, the results of CCEMG static and dynamic models, based on two setups of a LP weighted variable in accordance with the knowledge flows from (i), 23 OECD countries plus BRICs;[16] and (ii), all the OECD countries of the sample plus BRICs. Fourth, the results of the CCEMG models, based on a CH weighted foreign R&D variable in accordance with the three weighting configurations used for the LP R&D variable. Table 2 presents descriptive statistics for the variables. Here, the foreign R&D capital stock exhibits the highest average growth rate, whereas the TFP growth shows the lowest.

<<INSERT TABLE 2 HERE>>

The test that is implemented to analyze the cross-section dependence of residuals from the abovementioned models is the cross-section dependence (CD) test of Pesaran (2015). This study shows that the implicit null hypothesis of the CD test proposed by Pesaran (2004), who examines estimates of pair-wise error correlations, is the weak cross-section dependence of errors in the panel regression compared with the alternative of strong error cross-section dependence. Moreover, to investigate the stationarity of variables and the residuals from static models, we employ the second generation panel unit root test of Pesaran (2007) which allows for cross-section dependence across observations.[17] The null hypothesis for this test is that all panels contain unit roots across units, which is tested at a 5% level of significance. In general, this test yields unit root in all variables in levels.[18] We also provide the Root Mean Squared Error (RMSE) in the results.

## 6. Results

### 6.1. Static Econometric Models

Table 3 contains the results of the static models. Across models, the coefficients of domestic R&D are larger than those of the foreign R&D (except for the POLS and the CCEMG (i) estimates). More important, all the models with homogeneous slopes (except POLS) yield positive and statistically significant estimates of the domestic R&D at the 1% level, which range from 0.060 to 0.075, whereas the domestic R&D estimates from the MG and CDMG models vary between 0.039 and 0.061, all being statistically significant at the 10% level. Moreover, models which are restricted to homogeneous coefficients of the foreign R&D fall between 0.000 and 0.060, all being statistically significant at the 1% level except for the estimate from the first difference model. The MG and CDMG estimates of foreign R&D range from 0.025 to 0.031, where the foreign R&D estimate from the MG model is significant at the 10% level.

---

[16] Brazil, Russia, India and China.

[17] However, Pesaran et al. (2013) demonstrate that the Pesaran (2007) unit root test shows size distortions if there is more than one common factor. Consequently, it would be desirable in future empirical studies to implement either of the second generation unit root tests proposed by Pesaran et al. (2013), which have been designed to account for multiple unobserved common factors, even though no Stata routine has been developed so far: namely, the CIPS unit root test in the presence of multifactor error structure, or alternatively, the CSB Sargan-Bhargava test, augmented with cross-sectional averages, which has a better performance for smaller samples in $T$.

[18] Results of the implementation of this test on variables are presented in the online supplement. Results of this test on the residuals of static models are available upon request.

<<INSERT TABLE 3 HERE>>

On the basis of these results, we can state that even though the majority of the estimates of models where the interaction between unobservables is not modeled seem economically reliable, they are seriously misspecified for two reasons. First, the large CD statistic reflects a very high degree of residual cross-section dependence, so that the R&D estimates may be seriously biased and inconsistent (except for the POLS model); and second, all models have nonstationary residuals. Further, the CCEP model, which accounts for unobserved effects and is restricted to homogeneity in slopes, is also misspecified due to these two problems.

CCEMG estimates which account for heterogeneous technology parameters and the interaction between unobserved spillovers and shocks are also shown in Table 3 and follow two different setups: (i) a specification without a time trend; and (ii), a model in which a time trend is included. As can be seen, all coefficients of the domestic and foreign R&D variables are statistically significant and range from 0.054-0.090 to 0.057-0.061 respectively. The CCEMG models are not misspecified, since they have stationary and weak cross-sectional dependent residuals. Further, the estimates of the domestic and foreign R&D from the second CCEMG model and the foreign R&D coefficients from the first CCEMG model are larger than those in the misspecified pooled and MG models. As a result, we choose both CCEMG models; although the second CCEMG model is preferred because the former yields larger estimates and the RMSE is lower.

We also estimate CCEMG models which include spillover variables based on other weighting schemes. When we include LP foreign R&D variables which allow for knowledge dissemination from all OECD countries plus BRICs and from 23 OECD countries plus BRICs, the residuals have a low degree of cross-section dependence and are stationary, so that estimates are consistent and not seriously biased. The results are similar when a CH spillover variable that is based on the same weighting schemes is incorporated, so long as a time trend is excluded from the models. On the other hand, when a time trend is incorporated, specifications which include a CH foreign R&D variable with the three proposed weighting configurations are misspecified, due to strong residual cross-section dependence. Therefore, such estimates are seriously biased and inconsistent. The same happens when we include a CH variable which allows for knowledge diffusion from all the countries of the sample, and exclude a time trend (all these additional results are presented in section 3 of the online supplement).

**6.2. Dynamic Econometric Models**

Table 4 shows the results of the dynamic ARDL-POLS, 2FE and MG models where error cross-section independence is assumed and each model is estimated with $p = 1, 2$ and $3$ lags. The coefficients of domestic and foreign R&D from the dynamic POLS range from -0.013 to 0.008 and from -0.003 to 0.022 respectively, all being statistically insignificant. Estimates of the domestic and foreign R&D from the dynamic 2FE fall from -0114 to 0.005 and from 0.031 to 0.077, respectively, and are significant only for the specification with one lag. Meanwhile, the MG-ARDL estimates of the domestic R&D range from 0.025 to 0.060 and the coefficients of the foreign R&D fall from 0.007 to 0.024 where the domestic R&D coefficients are statistically significant for one and two lags. As can be seen, despite the fact that the variables are cointegrated at 1% in the long-term in all models, these dynamic models show a poor performance due to the strong cross-sectional dependence of residuals and, in the case of the ARDL-POLS and 2FE models, probably due to the erroneous pooling of the slopes of dynamic heterogeneous dynamic panels, as noted by Pesaran and Smith (1995). In consequence, none of the models has been chosen.

<<INSERT TABLE 4 HERE>>

The results of the CS-ARDL models, which regard the interaction between unobserved effects and include $p = 1, 2$ and 3 lags and a time trend, are shown in Table 5 column (i). Estimates of the domestic and foreign R&D variables range from 0.023 to 0.055 and from 0.070 to 0.083 respectively. Foreign R&D estimates are statistically significant at the 5% level, while the only domestic R&D estimate that is significant (at the 10% level) is that from the model with two lags. None of these models is misspecified, thanks to the fact that there is a low degree of residual cross-section correlation and the variables are cointegrated at 1% in the long-run. However, only the CS-ARDL specification which includes two lags obtains significant coefficients for both domestic and foreign R&D. It may be possible that the CS-ARDL models with one and three lags do not capture statistically significant domestic R&D estimates because of limitations on the time data, especially in the case of countries for which the data does not stretch beyond thirty years.

<<INSERT TABLE 5 HERE>>

The CS-ARDL (ii) model, which is a more flexible specification, has been estimated with one and two lags of variables and includes only two lagged cross-section averages and a time trend. As can be seen, the domestic and foreign R&D coefficients, which range from 0.066 to 0.085 and from 0.065 to 0.079 respectively, are significant at 5%. Moreover, these models are not misspecified, thanks to the low degree of the cross-section dependence of residuals and cointegration at 1%. These results indicate that if there were more observations or more flexibility in the addition of lags, then the CS-ARDL model might be more likely to yield positive and significant estimates. However, this flexibility has a cost, since the use of only two lagged cross-section averages may not be a suitable way to deal with the problem of reverse causality.

The results of the CS-DLMG models which account for unobservables and include one, two and three lags of the independent variables and a time trend suggest that all the domestic R&D estimates, which vary between 0.071 and 0.109, are significant at the 1% level, while the foreign R&D slopes are only significant at 1% for the specification with two and three lags and at 5% for one lag, falling from 0.052 to 0.080. Further, the reason why the CS-DLMG models are not misspecified is the low levels of residual cross-section dependence. Therefore, they do not yield seriously biased and inconsistent domestic R&D and foreign R&D estimates so long as feedback effects are not present. Although the RMSE of the CS-DLMG models is larger than that of the CS-ARDL models the Monte Carlo experiments in Tables 4 and 8 of Chudik et al. (2013) show that for samples lower than 100 cross-section and time observations, and in the absence of feedback effects, the CS-DLMG estimator is more efficient and has more power than the CS-ARDL model, even when the RMSE of the former is larger.

However, due to the characteristics of the R&D capital stock, we give priority to those dynamic models that account for feedback effects, although both the CS-ARDL and the CS-DLMG models yield complementary results. The CS-ARDL models indicate that it is possible to obtain consistent, positive, significant and not seriously biased estimates of domestic and foreign R&D, while the CS-DLMG models show that, with more complete data, these results may be more significant and the magnitude larger. Therefore, dynamic models which are not misspecified and yield significant estimates of both R&D variables are chosen. We can also see that (i), long-run cointegration is achieved at the 1% level across CS-ARDL models; (ii) the speed of cointegration is higher, compared to the traditional ARDL models; and (iii), the majority of significant domestic and foreign R&D estimates from Table 5 are more sizable than those from the standard ARDL models.

Similar results can be derived from models that include different weighing schemes for LP and CH spillover variables, except for those where a weighted CH foreign R&D variable with information on knowledge transmission from all countries has been incorporated. In that case, the CS-ARDL and CS-DLMG models are not as complementary as the models which include other spillover variables, insofar as most of the domestic R&D coefficients from the CS-ARDL models are insignificant and low (even for those which include 2 lagged cross-section averages), while the opposite happens with the CS-DLMG models. Therefore, under certain weighting configurations, the inclusion of the CH variable in the model may affect the domestic R&D estimates (the results are shown in section 4 of the online supplement).

## 7. Discussion

The empirical findings from the static and dynamic models reported in Tables 3 and 4 suggest that the CH framework ignores strong residual cross-section dependence due to unobserved effects, which may be correlated with the foreign and domestic R&D variables and lead to biased and inconsistent estimates. We can therefore infer that trade-related R&D spillovers cannot be estimated through the coefficient of a weighted foreign R&D variable whenever the CH approach does not account for the interaction between R&D spillovers transferred by any channel and other unobserved economic effects. In addition to these findings, Table 3 shows that a model that pools coefficients across countries and accounts for unobserved effects might not properly address a possible source of misspecification.

The results in Tables 3 and 5 and the online supplement show that several static and dynamic models which account for unobservables as sources of cross-section dependence and allow for heterogeneous technology coefficients yield favorable results for the weighted foreign R&D and domestic R&D estimates. However, that does not mean that those estimates represent large and genuine channel-specific R&D spillovers and returns to domestic R&D, respectively. Instead, as we anticipated in equation (8), those estimates embody a mixture of the effects of R&D variables and unobservables. This can easily be seen in the low degree of residual cross-section dependence which characterizes those estimates. The fact that, in most cases, domestic and foreign R&D estimates when unobserved effects are accounted for are more sizable compared to those of a CH specification corroborates these findings.

We can draw two conclusions from these results: first, when a weighted foreign R&D variable is intended to capture trade-related R&D spillovers in a situation where the interplay between R&D spillovers transferred by any channel and other unobserved common spillovers and shocks has been accounted for, its coefficient fails to capture pure channel-specific R&D spillovers because they cannot be separated from the effects of unobservables and at the same time be identified through the coefficient of the spillover regressor. In this case, the coefficient of a spillover variable might represent information about other factors which we cannot distinguish here. This problem is more pervasive for the estimates of the static models where a CH foreign R&D variable has been included, because the coefficient of this variable is subject to a high degree of residual cross-section dependence.

Second, the estimates of the domestic R&D variable yield a mix of returns to domestic R&D and the effect of spillovers and shocks associated with this variable, showing that they are not independent of each other, in contrast to the CH approach. However, the inclusion of a weighted foreign R&D variable and the effect of its associated shocks may have affected the magnitude of the domestic R&D estimates, as we predicted in equation (8). In fact, additional results of the static and dynamic CCEMG models, which incorporate a weighted CH foreign R&D variable which allows for knowledge diffusion from all the countries of the sample, show drastic changes in the domestic R&D estimates compared to the estimates of other models.

## 8. Conclusion

The approach of Coe and Helpman (1995), which has been widely employed to estimate channel-specific R&D spillovers and aggregate returns to domestic R&D in studies of international R&D spillovers, rests on two assumptions: (i) that the interplay between R&D spillovers transferred by any channel and other unobserved common spillovers and shocks does not bring about error cross-section dependence; and (ii) a weighted foreign R&D variable should be imposed to capture genuine channel-specific R&D spillovers, where it is assumed that these spillovers do not arise from unobservables, which remain neutral to TFP and R&D, but only arise from the spillover variable.

Yet, the CH framework does not clarify how a spillover variable can technically separate knowledge flows from other unobserved common effects, and how its coefficient can capture genuine channel-specific R&D spillovers: it forgets that this sort of effect might arise together with other unobserved effects common across countries as sources of cross-section dependence. The coefficient of a spillover variable in this case might represent other sorts of effects, different from pure R&D spillovers. By contrast, we have found that disregarding the interaction between R&D spillovers and other unobserved effects yields inconsistent domestic and weighted foreign R&D estimates since unobservables may be strongly cross-sectionally correlated with the variables of the model.

There is also no empirical justification for introducing a weighted foreign R&D variable when the simultaneous effect of unobserved spillovers and shocks is regarded. In the first place, since this variable cannot successfully separate R&D spillovers from unobservables, its coefficient fails to capture genuine channel-specific knowledge spillovers; and in the second, the estimates of the domestic R&D, which yield a mix of returns to domestic R&D and unobserved factors associated with domestic R&D, are affected by the unobserved effects which are associated with the weighted foreign R&D variable.

In accordance with these findings, the use of a weighted foreign R&D regressor might not be suitable for estimating channel-specific R&D spillovers within the abovementioned scenarios. Therefore, more research needs to be done on new approaches which estimate trade-related and channel-specific R&D spillovers in general in an empirical framework which would account for the interplay between this effect, R&D spillovers transferred by other channels, and other unobserved common macro and microeconomic spillovers and shocks associated with productivity and R&D. Identifying the effect of R&D spillovers from other unobserved effects is essential for drafting sound policies for R&D adoption in developing countries.

We agree with Keller (2010) when he writes about the importance of distinguishing the effects of knowledge spillovers from those of other possibly unobserved effects in analyzing technology spillovers from vertical FDI (which can easily be applied to trade and other channels of transmission of knowledge): *"it will be crucial to separate true technology spillovers from arms-length technology transactions, linkage effects, and measurement spillovers associated with vertical FDI, because the case for public policy intervention rests with the former, not the latter."*

TABLE 1

Sample description

| # | Country | TFP | Coverage | Rd | Coverage | Rf | Coverage | # | Country | TFP | Coverage | Rd | Coverage | Rf | Coverage |
|---|---|---|---|---|---|---|---|---|---|---|---|---|---|---|---|
| 1 | Argentina | 42 | 1970-2011 | 42 | 1970-2011 | 42 | 1970-2011 | 26 | Italy | 42 | 1970-2011 | 42 | 1970-2011 | 42 | 1970-2011 |
| 2 | Australia | 42 | 1970-2011 | 38 | 1973-2010 | 42 | 1970-2011 | 27 | Japan | 42 | 1970-2011 | 42 | 1970-2011 | 42 | 1970-2011 |
| 3 | Austria | 42 | 1970-2011 | 42 | 1970-2011 | 42 | 1970-2011 | 28 | Korea | 42 | 1970-2011 | 42 | 1970-2011 | 42 | 1970-2011 |
| 4 | Brazil | 42 | 1970-2011 | 38 | 1973-2010 | 42 | 1970-2011 | 29 | Malaysia | 42 | 1970-2011 | 24 | 1988-2011 | 42 | 1970-2011 |
| 5 | Bulgaria | 42 | 1970-2011 | 32 | 1980-2011 | 42 | 1970-2011 | 30 | Mexico | 42 | 1970-2011 | 42 | 1970-2011 | 42 | 1970-2011 |
| 6 | Canada | 42 | 1970-2011 | 42 | 1970-2011 | 42 | 1970-2011 | 31 | Netherlands | 42 | 1970-2011 | 42 | 1970-2011 | 42 | 1970-2011 |
| 7 | Chile | 42 | 1970-2011 | 32 | 1979-2010 | 42 | 1970-2011 | 32 | New Zealand | 42 | 1970-2011 | 40 | 1972-2011 | 42 | 1970-2011 |
| 8 | China | 42 | 1970-2011 | 24 | 1988-2011 | 42 | 1970-2011 | 33 | Norway | 42 | 1970-2011 | 42 | 1970-2011 | 42 | 1970-2011 |
| 9 | Colombia | 42 | 1970-2011 | 41 | 1971-2011 | 42 | 1970-2011 | 34 | Panama | 42 | 1970-2011 | 25 | 1986-2010 | 42 | 1970-2011 |
| 10 | Costa Rica | 42 | 1970-2011 | 38 | 1974-2011 | 42 | 1970-2011 | 35 | Peru | 42 | 1970-2011 | 34 | 1971-2004 | 42 | 1970-2011 |
| 11 | Cyprus | 42 | 1970-2011 | 32 | 1980-2011 | 42 | 1970-2011 | 36 | Philippines | 42 | 1970-2011 | 38 | 1970-2007 | 42 | 1970-2011 |
| 12 | Denmark | 42 | 1970-2011 | 39 | 1973-2011 | 42 | 1970-2011 | 37 | Poland | 42 | 1970-2011 | 27 | 1985-2011 | 42 | 1970-2011 |
| 13 | Ecuador | 42 | 1970-2011 | 39 | 1970-2008 | 42 | 1970-2011 | 38 | Portugal | 42 | 1970-2011 | 42 | 1970-2011 | 42 | 1970-2011 |
| 14 | Egypt | 42 | 1970-2011 | 39 | 1973-2011 | 42 | 1970-2011 | 39 | Romania | 24 | 1988-2011 | 23 | 1989-2011 | 42 | 1970-2011 |
| 15 | Estonia | 22 | 1990-2011 | 20 | 1992-2011 | 20 | 1992-2011 | 40 | Russia | 22 | 1990-2011 | 22 | 1990-2011 | 20 | 1992-2011 |
| 16 | Finland | 42 | 1970-2011 | 42 | 1970-2011 | 42 | 1970-2011 | 41 | Singapore | 42 | 1970-2011 | 42 | 1970-2011 | 42 | 1970-2011 |
| 17 | France | 42 | 1970-2011 | 42 | 1970-2011 | 42 | 1970-2011 | 42 | Spain | 42 | 1970-2011 | 42 | 1970-2011 | 42 | 1970-2011 |
| 18 | Germany | 42 | 1970-2011 | 41 | 1971-2011 | 42 | 1970-2011 | 43 | Sweden | 42 | 1970-2011 | 42 | 1970-2011 | 42 | 1970-2011 |
| 19 | Greece | 42 | 1970-2011 | 38 | 1970-2007 | 42 | 1970-2011 | 44 | Switzerland | 42 | 1970-2011 | 39 | 1970-2008 | 42 | 1970-2011 |
| 20 | Hungary | 42 | 1970-2011 | 38 | 1972-2009 | 42 | 1970-2011 | 45 | Thailand | 42 | 1970-2011 | 40 | 1970-2009 | 42 | 1970-2011 |
| 21 | Iceland | 42 | 1970-2011 | 40 | 1970-2009 | 42 | 1970-2011 | 46 | Turkey | 42 | 1970-2011 | 42 | 1970-2011 | 42 | 1970-2011 |
| 22 | India | 42 | 1970-2011 | 38 | 1970-2007 | 42 | 1970-2011 | 47 | United Kingdom | 42 | 1970-2011 | 42 | 1970-2011 | 42 | 1970-2011 |
| 23 | Indonesia | 42 | 1970-2011 | 38 | 1972-2009 | 42 | 1970-2011 | 48 | United States | 42 | 1970-2011 | 42 | 1970-2011 | 42 | 1970-2011 |
| 24 | Ireland | 42 | 1970-2011 | 42 | 1970-2011 | 42 | 1970-2011 | 49 | Uruguay | 42 | 1970-2011 | 41 | 1970-2010 | 42 | 1970-2011 |
| 25 | Israel | 42 | 1970-2011 | 42 | 1970-2011 | 42 | 1970-2011 | 50 | Venezuela | 42 | 1970-2011 | 31 | 1970-2000 | 42 | 1970-2011 |
|  |  |  |  |  |  |  |  |  | Total Obs | 2042 |  | 1873 |  | 2056 |  |

*Notes*: Variables: Log Total Factor Productivity (TFP), Log Domestic R&D (Rd) and Log Foreign R&D (Rf) capital stocks. All monetary variables are expressed in constant millions of US dollars of 2005 based on purchasing power parity (PPP). Definitions of these variables in the appendix.

TABLE 2

*Summary statistics*

| VARIABLES | Mean | Median | SD | Minimum | Maximum |
|---|---|---|---|---|---|
| Levels | | | | | |
| Total Factor Productivity | 0.96 | 0.97 | 0.14 | 0.57 | 1.60 |
| Domestic R&D Capital Stock (million PPP constant 2005 dollars) | 70858.70 | 9109.51 | 218128.50 | 48.66 | 2220345.00 |
| Foreign R&D Capital Stock (million PPP constant 2005 dollars) | 9325.21 | 3062.48 | 16401.86 | 4.45 | 174997.40 |
| Logarithms | | | | | |
| Log Total Factor Productivity | -0.05 | -0.03 | 0.14 | -0.56 | 0.47 |
| Log Domestic R&D Capital Stock (million PPP constant 2005 dollars) | 9.20 | 9.12 | 2.06 | 3.88 | 14.61 |
| Log Foreign R&D Capital Stock (million PPP constant 2005 dollars) | 7.99 | 8.03 | 1.63 | 1.49 | 12.07 |
| Growth | | | | | |
| Δ Total Factor Productivity | 0.00 | 0.01 | 0.03 | -0.25 | 0.19 |
| Δ Domestic R&D Capital Stock (million PPP constant 2005 dollars) | 0.04 | 0.04 | 0.06 | -0.16 | 0.34 |
| Δ Foreign R&D Capital Stock (million PPP constant 2005 dollars) | 0.07 | 0.06 | 0.17 | -1.09 | 3.09 |

*Notes*: These descriptive statistics refer to the sample of N = 50 countries from 1970 to 2011.

TABLE 3

*Static panel data models*

| Estimators | POLS | 2FE | FD | CCEP | MG | CDMG | CCEMG (i) | CCEMG (ii) |
|---|---|---|---|---|---|---|---|---|
| TFP dependent variable | | | | | | | | |
| Independent variables | | | | | | | | |
| Rd | -0.015*** | 0.075*** | 0.060*** | 0.071*** | 0.039* | 0.061* | 0.054** | 0.090*** |
| std errors | (0.003) | (0.005) | (0.015) | (0.009) | (0.020) | (0.036) | (0.023) | (0.021) |
| Rf | 0.021*** | 0.060*** | 0.000 | 0.037*** | 0.031* | 0.025 | 0.057*** | 0.061*** |
| std errors | (0.005) | (0.008) | (0.010) | (0.009) | (0.017) | (0.026) | (0.016) | (0.016) |
| CD-test | -0.28 | 119.82† | 183.45† | 28.1† | 13.00† | 3.63† | -0.21 | -0.59 |
| Order of Integration | I(1) | I(1) | I(1) | I(1) | I(1) | I(1) | I(0) | I(0) |
| RMSE | 0.142 | 0.094 | 0.029 | 0.045 | 0.058 | 0.066 | 0.035 | 0.032 |
| NXT | 1871 | 1871 | 1821 | 1871 | 1871 | 1871 | 1871 | 1871 |
| N | 50 | 50 | 50 | 50 | 50 | 50 | 50 | 50 |

*Notes:* log total factor productivity (TFP) is the dependent variable. log domestic R&D capital stock (Rd) and log foreign R&D capital stock defined by Lichtenberg and van Pottelsberghe de la Potterie (1998) (Rf) (allowing for knowledge diffusion from all countries of the sample) are the independent variables. A constant term is included but not reported. Estimators: 1) POLS Pooled OLS (augmented with T-1 year dummies). 2) 2FE: Two-way fixed effects (augmented with T-1 year dummies and N-1 country dummies). 3) FD: First Differences (augmented with T-2 year dummies because when differencing, a dummy for a year is dropped to avoid perfect multicollinearity). 4) CCEP: Pooled Pesaran (2006) augmented with common country dummies and cross-section averages, 5) MG: Mean Group. 6) CDMG: Cross-sectionally demeaned MG. 7) CMG: Common Correlated Effects MG Pesaran (2006) augmented with cross-section averages is presented in two versions: (i) without a time trend, and (ii) including a time trend. White heteroskedasticity-robust standard errors are reported in parentheses. Levels of significance are represented by * 10%, ** 5% and *** 1%. Diagnostics: (evaluated at the 5% level of significance, full results of the following tests are available on request): 1) CD test: The Pesaran (2015) test which is based on Pesaran (2004), for which Ho: Cross-section weak dependence of the residuals. 2) CIPS test: The Pesaran (2007) test evaluates the order of integration of the residuals where I(0): stationary, I(1): nonstationary. The root mean squared error (RMSE), NXT number of country-time observations and N number of countries are also included. † indicates that the null hypothesis of weak cross-section dependence of the residuals at the 5% level is rejected.

TABLE 4

*Dynamic ARDL panel data models assuming cross-section independence of errors, in a ECM representation*

| Estimators | POLS | | | 2FE | | | MG | | |
|---|---|---|---|---|---|---|---|---|---|
| | 1 lag | 2 lags | 3 lags | 1 lag | 2 lags | 3 lags | 1 lag | 2 lags | 3 lags |
| TFP dependent variable | | | | | | | | | |
| Independent variables | | | | | | | | | |
| Rd | -0.013 | -0.001 | 0.008 | -0.114*** | -0.015 | 0.005 | 0.025 | 0.059** | 0.060* |
| std errors | (0.011) | (0.011) | (0.010) | (0.043) | (0.036) | (0.036) | (0.030) | (0.029) | (0.032) |
| Rf | 0.022 | 0.007 | -0.003 | 0.077* | 0.053 | 0.031 | 0.024 | -0.004 | -0.007 |
| std errors | (0.014) | (0.014) | (0.013) | (0.045) | (0.037) | (0.034) | (0.028) | (0.029) | (0.031) |
| Cointegration coefficients | -0.058*** | -0.054*** | -0.056*** | -0.057*** | -0.069*** | -0.070*** | -0.235*** | -0.298*** | -0.345*** |
| std errors | (0.005) | (0.005) | (0.005) | (0.008) | (0.008) | (0.008) | (0.020) | (0.025) | (0.033) |
| CD-test | 156.35† | 122.15† | 122.02† | 148.78† | 115.62† | 117.05† | 19.26† | 16.52† | 14.49† |
| RMSE | 0.028 | 0.027 | 0.026 | 0.028 | 0.027 | 0.026 | 0.023 | 0.021 | 0.019 |
| NXT | 1821 | 1771 | 1721 | 1821 | 1771 | 1721 | 1821 | 1771 | 1721 |
| N | 50 | 50 | 50 | 50 | 50 | 50 | 50 | 50 | 50 |

*Notes*: log total factor productivity (TFP) is the dependent variable. log domestic R&D capital stock (Rd) and log foreign R&D capital stock defined by Lichtenberg and van Pottelsberghe de la Potterie (1998) (Rf) (allowing for knowledge flows from all the countries of the sample) are the independent variables. A constant term is included but not reported. Long run estimates and cointegration coefficients are reported. The estimators for autoregressive distributed lagged (ARDL) panel data specifications, which are represented by a Error Correction Model (ECM), are the following: 1) Dynamic ARDL POLS Pooled OLS (augmented with T-1 year dummies). 2) Dynamic ARDL 2FE: Two-way fixed effects (augmented with T-1 year dummies and N-1 country dummies). 3) Dynamic ARDL MG: Mean Group. White heteroskedasticity-robust standard errors are reported in parentheses. The POLS, 2FE and MG models are augmented with p=1, 2 and 3 lagged covariates. Levels of significance are represented by * 10%, ** 5% and *** 1%. Diagnostics: See Table 3, except for the CIPS test.

TABLE 5

*Dynamic panel data models accounting for cross-section dependence of errors, in a ECM representation*

| Estimators | CS-ARDL (ECM) | | | | | | CS-DLMG | | |
|---|---|---|---|---|---|---|---|---|---|
| | (i) | | | (ii) | | | | | |
| | 1 lag | 2 lags | 3 lags | 1 lag | 2 lags | | 1 lag | 2 lags | 3 lags |
| TFP dependent variable | | | | | | | | | |
| Independent variables | | | | | | | | | |
| Rd | 0.023 | 0.055* | 0.050 | 0.066** | 0.085** | | 0.071*** | 0.096*** | 0.109*** |
| std errors | (0.029) | (0.029) | (0.037) | (0.032) | (0.035) | | (0.018) | (0.028) | (0.035) |
| Rf | 0.083** | 0.070** | 0.082** | 0.079** | 0.065** | | 0.052** | 0.068*** | 0.080*** |
| std errors | (0.033) | (0.031) | (0.037) | (0.033) | (0.033) | | (0.021) | (0.024) | (0.028) |
| Cointegration coefficients | -0.436*** | -0.528*** | -0.626*** | -0.395*** | -0.469*** | | | | |
| std errors | (0.040) | (0.057) | (0.077) | (0.032) | (0.046) | | | | |
| CD-test | -1.61 | 0.35 | 0.70 | -1.34 | 0.34 | | -1.64 | -0.90 | -0.30 |
| RMSE | 0.013 | 0.011 | 0.013 | 0.015 | 0.013 | | 0.021 | 0.018 | 0.017 |
| NXT | 1720 | 1640 | 1579 | 1791 | 1735 | | 1758 | 1741 | 1687 |
| N | 48 | 45 | 43 | 50 | 48 | | 50 | 50 | 48 |

*Notes*: log total factor productivity (TFP) is the dependent variable. log domestic R&D capital stock (Rd) and log foreign R&D capital stock defined by Lichtenberg and van Pottelsberghe de la Potterie (1998) (Rf) (allowing for R&D transmission from all the countries of the sample) are the independent variables. A constant term is included but not reported. Long run estimates and cointegration coefficients are reported. The estimators for autoregressive distributed lagged (ARDL) panel data specifications, which are represented by a Error Correction Model (ECM), are the following: 1) Dynamic cross-sectional ARDL Chudik and Pesaran (2013a) (CS-ARDL-i) (augmented with three lags of the cross-sectional averages of the dependent and independent variables). 2) Dynamic cross-sectional ARDL (CS-ARDL-ii) (augmented with two lags of the cross-sectional averages of the dependent and independent variables). 3) Cross-sectional DL Chudik et al. (2013) Mean Group: CS-DLMG (augmented with three lags of the cross-sectional averages of the dependent and independent variables). White heteroskedasticity-robust standard errors are reported in parentheses. All models include a time trend. CS-ARDL (i) models are augmented with p=1, 2 and 3 lagged dependent and independent variables. CS-ARDL (ii) models are augmented with p=1 and 2 lags. CS-DLMG models are augmented with p=1, 2 and 3 lagged independent variables. Levels of significance are represented by * 10%, ** 5% and *** 1%. Diagnostics: See Table 3, except for the CIPS test.

**Appendix**

Data for TFP at constant national prices (2005=1) have been taken from the Penn World Table (PWT) 8.0 which, according to Inklaar and Timmer (2013), can be regarded as a measure of productivity growth in the following equation:

$$RTFP_{t,t-1}^{NA} \equiv \frac{RGDP_t^{NA}}{RGDP_{t-1}^{NA}} \Big/ Q_{t,t-1}^T \quad (A.1)$$

where *RTFP* and *RGDP* are the Total Factor Productivity and the GDP, respectively, both based on constant national prices. $Q_{t,t-1}^T$ is the Törnqvist quantity index of factor inputs. To construct RTFP, labor shares and depreciation rates of the capital stock vary across countries and over time. Further, initial capital stock starts from a capital/output ratio.

The Domestic R&D Capital Stock ($R_{it}^d$) is defined at constant PPPs of 2005 in millions of US dollars. This is constructed using the perpetual inventory method proposed by Klenow and Rodriguez-Clare (1997), where the initial observation starts in the same way as the capital/output ratio. This is as follows:

$$(R^d/Y)_{i0} = (Rex/Y)_i / (\delta^{Rd} + g_i) \quad (A.2)$$

where $(R^d/Y)_{i0}$ is the ratio of the domestic R&D capital stock to GDP in the initial period 0 in country $i$, $(Rex/Y)$ is the average Gross Expenditure on R&D (GERD) to GDP, divided by the sum of (i), the domestic R&D capital stock rate of depreciation $\delta^{Rd}$, which is set as 0.15, following Griliches (1998); and (ii), an estimate of the average growth rate of the GDP of country $i$ from 1981-1990 $g_i$ (for a country whose GDP series begins in 1990, the average growth is measured by starting at some point between 1990 and 2000). To find the initial domestic R&D capital stock, the equation (A.2) is multiplied by the initial GDP. Next, the following equation is used to complete the rest of the series:

$$R_{it}^d = (1 - \delta^R) R_{i,t-1}^d + Rex_{it} \quad (A.3)$$

where $R_{it}^d$ is the domestic R&D capital stock and $Rex_{it}$ the GERD.

To construct these series, I take data on GERD as a percentage of GDP from four different sources in the following order: (i) The UNESCO Institute for Statistics on Science, Technology and Innovation Database from 1996-2010. (ii) The UNESCO Statistical Yearbook (1999) from 1980-1995 (and for some countries to 1996). This source defines GERD as a percentage of GNP. Therefore, to convert it to a percentage of GDP, it has been multiplied by the Ratio of GNP to GDP (divided by 100) from the PWT 7.1. (iii), The OECD Main Science and Technology Indicators Statistics database from 1980-2011. (iv) Lederman and Saenz (2005), which includes information on GERD as a percentage of GDP from different series of the UNESCO Statistical Yearbook. I take data from this source from 1970-2005.

In the case of Finland, Greece, Iceland, Ireland, Portugal, Singapore, Sweden, Thailand, the United Kingdom and Uruguay, we have taken data for the period before 1970 from the fourth source to linearly interpolate them with post-1970 data to complete the data series from 1970 onwards. Once this was done, pre-1970 observations were dropped. The data collection is summarized in Tables D1 and D2 in the online supplement. Missing data have been linearly interpolated according to the data availability of each country. Initial data on GERD as a percentage of GDP were used to obtain the first observations for Domestic R&D capital stock. I multiplied this by the output-side real GDP at chained 2005 PPPs in millions of US dollars, a measure of the production possibilities of an economy, from the PWT 8.0. With this I obtained the PPP Converted Expenditure on R&D (GERD) at 2005 constant prices in millions of US dollars, and I used it to construct the rest of the Domestic R&D capital stock series.

The weighted Foreign R&D Capital Stock ($R^f$) is defined by Lichtenberg and van Pottelsberghe de la Potterie (1998) as follows:

$$R_{it}^f = \sum_{i \neq c} (M_{ic}/Y_c)_t \, R_{ct}^d \tag{A.4}$$

where $M_{ic}$ is country $i$'s imports of goods and services from country $c$, $Y_c$ is the GDP in country $c$ and $R_{ct}^d$ is the domestic R&D capital stock. Data for $M_{ic}$ were taken from the bilateral imports on a c.i.f. basis in US current dollars from the IMF Direction of Trade Statistics (DOTS). To get data for $Y_c$, I multiplied the GDP at current national prices in local currency times the exchange rate of national currency per USD at the market value, both from the PWT8.0. As a result, the foreign R&D capital stock is defined at constant PPPs of 2005 in millions of US dollars.

Foreign R&D capital stock is also measured according to Coe and Helpman (1995) as follows:

$$R_{it}^{f-CH} = \sum_{i \neq c} w_{ic,t} \, R_{ct}^d \tag{A.5}$$

where $w_{ic,t} = (M_{ic}/\sum_{i \neq c} M_{ic})_t$ and $\sum_{i \neq c} w_{ic,t} = 1$.